# Analytic calculation of transition dipole moment using four-component relativistic equation-of-motion coupled-cluster expectation value approach

Tamoghna Mukhopadhyay[1], Sudipta Chakraborty[1], Somesh Chamoli[1], Malaya K. Nayak[2,3] [†] and Achintya Kumar Dutta[1] [*]

1. Department of Chemistry, Indian Institute of Technology Bombay, Powai, Mumbai-400076, India.
2. Theoretical Chemistry Section, Bhabha Atomic Research Centre, Trombay, Mumbai 400085, India.
3. Homi Bhabha National Institute, BARC Training School Complex, Anushakti Nagar, Mumbai 400094, India.

## Abstract

We have developed an efficient scheme for the calculation of transition properties within the four-component relativistic equation-of-motion coupled cluster (EOM-CC) method using the expectation value approach. The calculation of transition properties within the relativistic EOM-CC framework requires the solution of both right and left eigenvectors using Davidson's iterative diagonalization scheme. The accuracy of the approach has been investigated by calculating low-lying transitions of Xe atom, HI molecule and spin forbidden $^1S_0 \rightarrow {}^3P_1$ and spin allowed $^1S_0 \rightarrow {}^1P_1$ transitions in few closed shell cations. Additionally, applications aimed at evaluating the L-edge X-ray absorption spectrum (XAS) of Ar atom is studied. The calculated results show good agreement with the earlier theoretical studies and experimental values.

[†] mknayak@barc.gov.in; mk.nayak72@gmail.com

[*] achintya@chem.iitb.ac.in

# 1. INTRODUCTION

The study of excited state properties has gained considerable attention in theoretical chemistry and quantum physics. The ability to accurately compute transition dipole moments is indispensable for predicting various spectroscopic observables, such as absorption and emission spectra, as well as investigating photochemical processes, energy transfer mechanisms, and electronic structure in complex molecular systems. Among the various electronic structure methods available, the single reference coupled cluster (CC) method[1] has emerged as one of the most popular ones due to its black box nature and capability to include the electron correlation effect systematically. When there are Coulomb interactions between electrons in atoms and molecules, the size-extensive CC-based formulation works especially well. The singles and doubles (CCSD) variant of CC-based methods has shown to be very effective in yielding results that are reasonably good. One can capture the more accurate effect of correlation and systematically reach the full configuration interaction (FCI) limit, with the size extensivity of the energy guaranteed even at the truncated level of the cluster operator. The ground state coupled cluster method is extended to the excited using the equation-of-motion (EOM) approach[2]. The equation of motion coupled cluster (EOM-CC) method gives identical results to that of the coupled cluster linear response (CC-LR) approach for excitation energy calculations[3–5]. Although both approaches have completely different theoretical origins and CCLR provides a size-intensive transition dipole moment (TDM)[6], whereas EOM-CCSD fails to maintain a size-intensive nature in this scenario. However, the transition moment in EOM-CCSD is generally in good agreement with the LR-CCSD method in most of the cases[7].

One needs to account for relativistic effects for the simulation of heavy-element systems[8]. The relativistic formalism can be constructed either by using a four-component(4c) Dirac-Coulomb (DC) Hamiltonian in conjunction with the "no pair approximation" to get rid of negative energy states or the two-component Hamiltonian generated by block-diagonalizing the 4c Hamiltonian to decouple the electronic and positronic states[9]. The EOM-CCSD method based on 4c DC Hamiltonian can accurately include both relativistic and electronic correlation effects[10–12]. The four-component relativistic EOM-CC (4c-EOM-CC) is generally used in the singles and doubles approximation (4c-EOM-CCSD), and its $O(N^6)$ power of the basis set, which restricts its use beyond small molecules. It should be noted that the computational cost of 4c-EOM-CCSD is at least thirty-two times higher than the corresponding non-relativistic version[9]. Various strategies, from perturbative approximation[13] and density fitting[14] to natural spinors[15], can be used to reduce the computational cost of the 4c-EOM-CCSD method. The most salient characteristic of heavy elements is the spin-orbit coupling, which leads to intensity borrowing by spin-forbidden transitions, fine structure splitting, and many other interesting phenomena. One needs to calculate transition properties, in addition to excitation energies, to simulate the spectra of these systems. The theory of relativistic EOM-CC methods is quite well developed and generally available for energy calculations[12]. However, accurate simulation of experimental observable requires the calculation of transition properties within the 4c-EOM-CCSD method. Oleynichenko and co-workers[16,17] have reported the transition properties calculation within the Fock space relativistic coupled cluster[18,19] framework using a finite-field approach. The aim of this manuscript is to describe the theory, implementation, and benchmarking of analytical computation of transition properties within the framework of the 4c-EOM-CCSD method.

## 2. THEORY

### 2.1 Relativistic Coupled Cluster method:

The relativistic coupled cluster wave function is defined as

$$|\Psi_{CC}\rangle = e^{\hat{T}}|\Phi_0\rangle \quad (1)$$

Here, the wavefunction $|\Phi_0\rangle$ is obtained from the solution of the Dirac-Hatree-Fock (DHF) equation using a 4c-DC Hamiltonian, whose form is given as

$$\begin{bmatrix} \hat{V} + \hat{J} - \hat{K} & c(\sigma.\hat{P}) - \hat{K} \\ c(\sigma.\hat{P}) - \hat{K} & \hat{V} - 2mc^2 + \hat{J} - \hat{K} \end{bmatrix} \begin{bmatrix} \Phi^L \\ \Phi^S \end{bmatrix} = E \begin{bmatrix} \Phi^L \\ \Phi^S \end{bmatrix} \quad (2)$$

where $|\Phi^L\rangle$ and $|\Phi^S\rangle$ denote large and small components of 4-spinor $|\Phi_0\rangle$. The $\hat{V}$ in Eq. (2) denotes electro-nuclear interaction. Here, $\hat{P}$ represents the momentum operator, $\sigma$ denote Pauli spin matrices, $m$ is the mass of the electron, and $c$ is the speed of light. The direct electron-electron interaction is $\hat{J}$, and the exchange operator is denoted by $\hat{K}$.

The $\hat{T}$ in Eq. (1) is the cluster operator defined as:

$$\hat{T} = \hat{T}_1 + \hat{T}_2 + \hat{T}_3 + \ldots \hat{T}_N \quad (3)$$

where

$$\hat{T}_1 = \sum_{ia} t_i^a \{\hat{a}_a^\dagger \hat{a}_i\}$$
$$\hat{T}_2 = \frac{1}{4} \sum_{ijab} t_{ij}^{ab} \{\hat{a}_a^\dagger \hat{a}_b^\dagger \hat{a}_i \hat{a}_j\} \quad (4)$$

are one-electron ($\hat{T}_1$) and two-electron ($\hat{T}_2$) cluster operators written in normal-ordered formalism. In general, the $N$-spinor cluster operator $\hat{T}_N$ will have the form:

$$\hat{T}_N = \left(\frac{1}{N}\right)^2 \sum_{\substack{ij\ldots \\ ab\ldots}} t_{ij\ldots}^{ab\ldots} \{\hat{a}_a^\dagger \hat{a}_b^\dagger \ldots \hat{a}_i \hat{a}_j \ldots\} \quad (5)$$

where $t_i^a, t_{ij}^{ab}, \ldots t_{ij\ldots}^{ab\ldots}$ are the cluster amplitudes for the respective one, two, … $N$-electron cluster operators. The indices $i, j, k, l, \ldots$ and $a, b, c, d, \ldots$ represent occupied and virtual spinors, respectively. The relativistic coupled cluster method is generally used in singles and doubles truncation of the cluster a(CCSD), i.e.,

$$\hat{T} = \hat{T}_1 + \hat{T}_2 \quad (6)$$

which scales as $O(N^6)$ power of the basis set. Extension to triples and quadruples correction has also been achieved in the literature[20].

In the present case, we are using 4c-DC Hamiltonian ($H^{DCH}$) with no-pair approximation, and the coupled cluster similarity transformed Hamiltonian can be written as:

$$\bar{H}^{DCH} = e^{-\hat{T}} \hat{H}^{DCH} e^{\hat{T}} \qquad (7)$$

The coupled cluster energy and amplitudes can be obtained as follows:

$$\langle \Phi_0 | \bar{H}^{DCH} | \Phi_0 \rangle = E \qquad (8)$$

$$\langle \Phi_{ij...}^{ab...} | \bar{H}^{DCH} | \Phi_0 \rangle = 0 \qquad (9)$$

**2.2 Equation-of-motion formalism for excited state:**

In the equation-of-motion theory[2,5,21], the access to the target state ($|\Psi_k\rangle$) wave function can be obtained by the action of a linear excitation operator on the reference state ($|\Psi_0\rangle$) wave function:

$$|\Psi_k\rangle = \hat{R}_k |\Psi_{CC}\rangle \qquad (10)$$

The form of the linear excitation operator depends upon the nature of the target state, and for the excited state, it has a form of

$$\hat{R}_k = r_0 + \sum_{ia} r_i^a \{\hat{a}_a^\dagger \hat{a}_i\} + \sum_{i<j,a<b} r_{ij}^{ab} \{\hat{a}_a^\dagger \hat{a}_i \hat{a}_b^\dagger \hat{a}_j\} + ... \qquad (11)$$

The excited state Schrodinger equation for the $k^{th}$ excited state in the relativistic EOM-CC framework can be written as

$$\hat{H}_N^{DCH} \hat{R}_k e^{\hat{T}} |\Phi_0\rangle = E_k \hat{R}_k e^{\hat{T}} |\Phi_0\rangle \qquad (12)$$

One can directly calculate the excitation energy ($\omega_k = E_K - E_0$) using the commutator form of the Eq. (12).

$$[\bar{H}_N^{DCH}, \hat{R}_k] |\Phi_0\rangle = \omega_k \hat{R}_k |\Phi_0\rangle \qquad (13)$$

Being a non-Hermitian formalism, $\bar{H}_N^{DCH}$ also has a left eigenvector

$$\hat{L}_k = l_0 + \sum_{ia} l_a^i \{\hat{a}_i^\dagger \hat{a}_a\} + \sum_{i<j,a<b} l_{ab}^{ij} \{\hat{a}_i^\dagger \hat{a}_a \hat{a}_j^\dagger \hat{a}_b\} + ... \qquad (14)$$

with the same eigenvalues satisfying

$$\langle \Phi_{CC} | \hat{L}_k \bar{H}_N^{DCH} = \langle \Phi_{CC} | \hat{L}_k E_k \qquad (15)$$

These two sets of eigenvectors together satisfy a biorthogonal condition,

$$\langle \Phi_{CC} | \hat{L}_k \hat{R}_k | \Phi_{CC} \rangle = \delta_{kl} \qquad (16)$$

**2.3 Transition dipole moment using expectation value approach:**

The transition properties in the EOM-CCSD framework are calculated as an expectation value[21]. Due to the non-hermitian nature of the coupled cluster similarity transformed Hamiltonian, the ket states are not the same as the complex conjugate of the bra states. For the property calculations, one needs to construct both right and left eigenvectors. The corresponding ket states of $|\Psi_{CC}\rangle$ and $|\Psi_k\rangle$ are

$$\langle \tilde{\Psi}_{CC}| = \langle \Phi_0|(1+\hat{\Lambda})e^{-\hat{T}} \tag{17}$$

$$\langle \tilde{\Psi}_k| = \langle \Phi_0|\hat{L}_k e^{-\hat{T}} \tag{18}$$

respectively, where $\hat{L}_k$ is the EOM-CC left eigenvector defined in Eq. (14) and $\hat{\Lambda}$ is the coupled cluster de-excitation operator[1] represented as

$$\hat{\Lambda} = \hat{\Lambda}_1 + \hat{\Lambda}_2 + \hat{\Lambda}_3 + ... \tag{19}$$

where

$$\hat{\Lambda}_1 = \sum_{ia} \lambda_a^i \{\hat{a}_i^\dagger \hat{a}_a\}$$
$$\hat{\Lambda}_2 = \frac{1}{4}\sum_{ijab} \lambda_{ab}^{ij} \{\hat{a}_i^\dagger \hat{a}_a \hat{a}_j^\dagger \hat{a}_b\} \tag{20}$$

and in general

$$\hat{\Lambda}_n = \left(\frac{1}{n!}\right)^2 \sum_{\substack{ij... \\ ab...}} \lambda_{ab...}^{ij...} \{\hat{a}_i^\dagger \hat{a}_a \hat{a}_j^\dagger \hat{a}_b ...\} \tag{21}$$

Now any first-order property can be calculated as the expectation value of the corresponding operator $\Theta$ as

$$\langle \Theta \rangle = \langle \Psi|\Theta|\Psi\rangle \tag{22}$$

Following Eq. (22) by analogy, we can write the square of the transition dipole moment ($\mu$) for the transition from $i^{th}$ to $k^{th}$ state

$$|\mu_{i\to k}|^2 = \langle \Phi_0|\hat{L}_i \bar{\mu} \hat{R}_k|\Phi_0\rangle \langle \Phi_0|\hat{L}_k \bar{\mu} \hat{R}_i|\Phi_0\rangle \tag{23}$$

where,

$$\bar{\mu} = e^{-\hat{T}}\mu e^{\hat{T}} \tag{24}$$

For the ground to excited state transition moment Eq. (23) can be expressed as

$$|\mu_{o\to k}|^2 = \langle \Phi_0|(1+\hat{\Lambda})\bar{\mu}\hat{R}_k|\Phi_0\rangle \langle \Phi_0|\hat{L}_k \bar{\mu}|\Phi_0\rangle \tag{25}$$

The quantity $|\mu_{i\to k}|^2$ can be expressed in terms of left and right transition moment

$$\mu_{0\to k} = \langle \Phi_0|(1+\hat{\Lambda})\bar{\mu}\hat{R}_k|\Phi_0\rangle \tag{26}$$
$$\mu_{k\to 0} = \langle \Phi_0|\hat{L}_k \bar{\mu}|\Phi_0\rangle \tag{27}$$

The left and right transition moment individually cannot be related to any observable as they are not individually normalized. However, the biorthogonality relation

$$\langle \Phi_0 | (1+\hat{\Lambda})\hat{R}_k | \Phi_0 \rangle = \langle \Phi_0 | \hat{L}_k | \Phi_0 \rangle = 0 \tag{28}$$

ensures proper normalization of the calculated $|\mu_{0 \to k}|^2$. It should be noted that the transition dipole moments are not an experimentally observable quantity. Instead, one can calculate the experimentally measurable quantity known as the oscillator strength ($f$).

$$f_{0 \to k} = \frac{2}{3} \Delta E_{0 \to k} |\mu_{0 \to k}|^2 \tag{29}$$

The coupled cluster linear response formalism provides an alternative method to calculate excitation energy and transition properties. It leads to identical expressions as the EOM-CC method for excitation energy and right transition moment, although the two approaches use completely different philosophies for describing the excited states[3]. The left transition moment, on the other hand, includes the response of the ground state coupled cluster amplitudes.

One can circumvent the explicit calculation of amplitude response by solving an additional set of equations for perturbation-independent parameters $\xi_k$ as

$$\langle \Phi_0 | (1+\hat{\Lambda})[\bar{H}, \hat{a}_p^\dagger \hat{a}_q] \hat{R}_k | \Phi_0 \rangle + \langle \Phi_0 | \xi_k [\bar{H}, \hat{a}_p^\dagger \hat{a}_q] \hat{R}_k | \Phi_0 \rangle + \omega_k \langle \Phi_0 | \xi_k \hat{a}_p^\dagger \hat{a}_q | \Phi_0 \rangle = 0 \tag{30}$$

Koch et al[3]. has demonstrated that both linear response and EOM-CC expectation value approach lead to identical results at the full configuration interaction (full-CI) limit. While preparing this manuscript, we came across an implementation of the linear response approach for ground state second order properties within a four-component relativistic coupled cluster framework by Gomes and co-workers[22]. It should be noted that the transition moments calculated using the EOM-CC expectation value approach is not size-intensive, except for the full-CI limit. The coupled cluster linear response transition moments are, on the other hand, size-intensive even at the truncated level of the cluster operator, although computationally, it is more expensive than the EOM-CC expectation value approach. In the majority of the cases, the EOM-CCSD transition moments are very close to the corresponding linear response results[7].

The relativistic EOM-CCSD transition moments can be efficiently calculated using a one-body reduced density matrix

$$(d)_{qp} = \langle \hat{a}_p^\dagger \hat{a}_q \rangle = \langle \Phi_0 | \hat{L}_k \hat{a}_p^\dagger \hat{a}_q \hat{R}_k | \Phi_0 \rangle \tag{31}$$

and the expectation value of transition dipole moment considering the transition from $i^{th}$ to any $k^{th}$ state in terms of reduced density matrix can be represented as

$$\mu_{ik} = \langle \Phi_0 | \hat{L}_i \bar{\mu} \hat{R}_k | \Phi_0 \rangle = Tr(\mu d_{i \to k}) \tag{32}$$

The relativistic EOM-CCSD transition moment, as described above, is implemented in our in-house quantum chemistry software package BAGH[23].

# 3. RESULTS AND DISCUSSION

We have tested the accuracy of the 4c-EOM-CCSD expectation value approach by calculating the transition properties corresponding to the low lying excited states and the corresponding transition dipole moments for the Xenon (Xe) atom, HI molecule, few closed shell cations, and Argon (Ar) atom. All the calculations are performed using our in-house software package BAGH[23] The Fock matrix, one and two-electron integrals in molecular spinor basis, and the dipole moments integrals are calculated using DIRAC[24]. For the Xe atom and HI molecule, we utilized the d-augment-dyall.aexz (x=2, and 3) basis, for the closed shell cations contracted and uncontracted , 6-31G and Sapproro-DZP-2012-ALL basis set were used, whereas for the Argon (Ar) atom fully uncontracted version of 6-311(2+, +) G(p,d) basis set was used. All the electrons were correlated for the calculations.

## 3.1. Xe Atom:

We employ the 4c-EOM-CCSD expectation value approach to compute the transition dipole moments of the first bright states Xe atom for lower-lying excited states. Table I compares the excitation energies and transition dipole moment(TDM) obtained from the 4c-EOM-CCSD method with the experiment values[25] and the earlier Fock-space relativistic coupled cluster (FSCC) results[16]. At the d-aug-dyall.ae2z basis set, the excitation energy in EOM-CCSD reasonably agrees with the corresponding experimental value for the first three bright states. However, the excitation energy corresponding to $5p^5$ $(^2P_{3/2})5d$ $^2[3/2]_1^0$ state is overestimated by more than 4000 cm$^{-1}$. The TDM values for the first two bright states show good agreement with the experimental value. Whereas the TDM corresponding to the $5p^5$ $(^2P_{3/2})5d$ $^2[1/2]_1^0$ and $5p^5$ $(^2P_{3/2})5d$ $^2[3/2]_1^0$ states are underestimated and overestimated with respect to the experimental values, respectively.

As we go from d-aug-dyall.ae2z to d-aug-dyall.ae3z basis set, the agreement of the excitation energy with the experimental value for all four states improves. The first two bright states undergo a red shift on going to the d-aug-dyall.ae3z basis set. The excitation energy corresponding to the state $5p^5$ $(^2P_{3/2})5d$ $^2[1/2]_1^0$ and $5p^5$ $(^2P_{3/2})5d$ $^2[3/2]_1^0$ states, on the other hand, undergo a blue shift. The error in the $5p^5$ $(^2P_{3/2})5d$ $^2[3/2]_1^0$ state decreases to 2439 cm$^{-1}$. The behavior of TDM, with the increase in basis set, does not follow a fixed pattern. The TDM values for the $5p^5$ $(^2P_{3/2})6s$ $^2[1/2]_1^0$ and $5p^5$ $(^2P_{3/2})6s$ $^2[3/2]_1^0$ remain almost unchanged on increasing the basis set. The TDM for the $5p^5$ $(^2P_{3/2})5d$ $^2[3/2]_1^0$ slightly increases, leading to a deteriorating agreement with the experiment. The only improvement is observed in the TDM for the $5p^5$ $(^2P_{3/2})5d$ $^2[1/2]_1^0$ state, where an increase in the basis set reduced the error with respect to the experiment to half. The agreement of the 4c-EOM-CCSD with the experiment is slightly inferior to that observed in FSCCSD[16].

## 3.2. Hydrogen Iodide (HI) Molecule:

The exploration of excited states of HI is highly intriguing and has been extensively investigated experimentally[26–30]. However, there remains a notable scarcity of accurate qualitative theoretical studies on these excited states. The existing literarature includes relativistic effective core potential-based configuration-interaction (CI) calculations by Chapman et al.[31–33], spin-orbit coupled CI calculations by Buenker and coworkers[34], and an empirical model-based analysis by Brown and coworkers[35]. The 4c-EOM-CCSD expectation value approach can help to address the lack of accurate theoretical data on the excited states of the HI molecule. Table II shows the excitation energies and the corresponding transition dipole moment values for the transition from the ground state (X $^1\Sigma^+$) to different low-lying excited states (a $^3\Pi_1$, a $^3\Pi_{0+}$, and A $^1\Pi_1$) of the HI molecule along with the previous theoretical[34] and experimental[36] estimates. The 4c-EOM-CCSD method shows excellent agreement with the previously reported MRD-CI values of Buenker and coworkers[34] for both excitation energy and transition moment. However, a small difference can be observed that the a $^3\Pi_{0+}$, and A $^1\Pi_1$ states have almost identical TDM in MRD-CI method, whereas the state a $^3\Pi_{0+}$ shows a slightly higher transition moment in 4c-EOM-CCSD. The 4c-EOM-CCSD excitation energy shows a good agreement with the experiment for the a $^3\Pi_1$ and and a $^3\Pi_{0+}$ state. However, the excitation energy corresoonding to the A$^1\Pi_1$ state is underestimated by 1592 cm-1 with respect to the experimental reference. There is no experimental value for the transition dipole moment available for comparison.

On increasing the basis set from d-aug-dyall.ae2z to d-aug-dyall.ae3z, the excitation energy values and TDM values for all the states increase. This led to a deterioration of the agreement with the experiment for the $^3\Pi_{0+}$ state, where the 4c-EOM-CCSD overestimates the excitation energy by 552 cm$^{-1}$. The error in the excitation energy for the A$^1\Pi_1$ slightly decreases on going to d-aug-all.ae3z basis set. However, it still displays an error of 1375 cm$^{-1}$ with respect to the experiment. A better agreement with the experiment will presumably needs triples correction to the EOM-CCSD method. The difference between the TDM of a $^3\Pi_{0+}$, and A $^1\Pi_1$ states also slightly decreases on going to d-aug-dyall.ae3z basis set

### 3.3. Absorption spectra of closed-shell cations:

The lowest optically active transition in closed shell cations (such as Na$^+$, Ca$^{2+}$ etc.) in the non-relativisitic framework is due to $^1S \rightarrow {}^1P$ transition. As a result, the absorption spectrum corresponding to the low-lying excited states is expected to display a single peak corresponding to this particular transition. In the context of the relativistic framework, in addition to the aforementioned peak ($^1S_0 \rightarrow {}^1P_1$), an additional, less intense peak is observed in the absorption spectra. The occurrence of which can be attributed to the effects of spin-orbit coupling, which causes the $^1S_0 \rightarrow {}^3P_0$ transition to become optically allowed.

Our objective is to assess the accuracy of the relativistic 4c-EOM-CCSD method by computing the splitting of peaks in closed shell cations and comparing it with the experimental estimate[37]. To achieve this goal, we have determined the excitation energies and transition dipole moments for the spin allowed ($^1S_0 \rightarrow {}^1P_1$) and spin-forbidden transitions ($^1S_0 \rightarrow {}^3P_0$) in few closed shell cations. Table III presents a comparison of excitation energies and transition dipole moments in different basis sets along with the experimental values and previously reported theoretical data. In the uncontracted 6-31G basis set, the calculated peaks associated with the $^1S_0 \rightarrow {}^3P_1$ and $^1S_0 \rightarrow {}^1P_1$ transitions in all the cations are found to be slightly underestimated compared to the experimental results. However, the results are in much better agreement with the experiment as compared to the two-component time-dependent equation of motion coupled cluster (X2C-TD-EOM-CCSD) results by Li and coworkers[38]. The latter results are significantly overestimated as compared to the experimental values with a mean absolute error(MAE) of 1.4061 eV. Our 4c-EOM-CCSD method shows an MAE of 0.3629 eV and also gives better agreement with the experimental splitting between the $^3P_0$ and $^1P_1$ states. The transition dipole moment values corresponding to the $^1S_0 \rightarrow {}^3P_1$ and $^1S_0 \rightarrow {}^1P_1$ transitions clearly validate the reliability of our method as it accurately predicts a lower transition moment value for less intense spin-forbidden transition and a high moment value for the spin-allowed transition. The poor performance of X2C-TD-EOM-CCSD[38] is presumably due to the small contracted 6-31G basis set, whereas an uncontracted version of the basis set is used in the present study.

The performance of the X2C-TD-EOM-CCSD[38] method significantly improves on going from the 6-31G to Sapproro-DZP-2012-ALL basis set. The excitation energy shows a significantly improved agreement with the experimental value. The MAE value of X2C-TD-EOM-CCSD excitation energy gets reduced to 0.0387 eV. Our 4c-EOM-CCSD method gives a comparable performance for the excitation energy and shows a slightly better performance in predicting the splitting. There is no pronounced change in transition dipole moment value for $Na^+$, $Mg^{2+}$, and $Ca^{2+}$ cations on increasing the basis set. However, the TDM value slightly increases for $K^+$ atom. The method gives significantly better performance than the relativistic time-dependent density functional (X2C-TDDFT) calculations[39,40].

To further validate the suitability of 4c-EOM-CCSD method for simulation of spin forbidden transitions, we have simulated the oscillator strengths corresponding to the low-lying excited states of $Na^+$ ion. Figure 1 shows the atomic absorption spectra generated through both four-component relativistic and non-relativistic EOM-CCSD methods in an uncontracted 6-31G basis set. The oscillator strength values of the associated transitions are also shown in thFigure 1. A single peak corresponding to the $^1S \rightarrow {}^1P$ transition is observed in the non-relativistic EOM-CCSD method. On the other hand, two distinct absorption peaks are observed in the relativistic case. In the relativistic domain, the spin-forbidden $^1S_0 \rightarrow {}^3P_1$ transition borrows intensity through spin-orbit coupling, which results in a small oscillator strength value. In contrast, the spin-allowed $^1S_0 \rightarrow {}^1P_1$ transition displays a substantial oscillator strength value relative to the former, indicating a significant transition probability.

### 3.4. X-ray absorption spectra of Argon atom:

In the previous sections, we have dealt with the excitation energies and transition moments of Xe and ions of alkaline metals originating from the valance spinors in the relativistic framework. X-ray absorption spectroscopy (XAS) has emerged as an effective tool to determine the local electronic structure of atoms and molecules[41,42]. The singly excited state from the core-orbital in EOM-CCSD remains embedded in the continuum of the doubly excited states from the valence orbital, which leads to convergence issues in the Davidson iterative diagonlization procedure[43]. One can bypass this problem by using the core valence separation (CVS) approximation of Cederbaum and co-workers[44]. The CVS-EOM-CCSD method is known for its accuracy in predicting XAS and XPS experimental spectra[43,45,46].

Most of the CVS-EOM-CCSD based simulations of of core-level spectroscopy are performed the non-relativistic picture. Recently, Gomes and co-workers have reported a CVS-EOM-CCSD method based on a 4c Dirac Coulomb Hamiltonian for excitation and ionization potential corresponding to core-level spectroscopy. Vidal et. al[47] have reported a frozen-core-CVS-EOM-CCSD method with perturbative inclusion of the spin-orbit coupling for the L-edge XAS absorption spectra. The L-edge spectra originates due to transition from the 2s or 2p orbital of an atom. Due to spin-orbital coupling (SOC), the 2p orbital can split into lower energy $^2P_{1/2}$ and higher energy $^2P_{3/2}$ states, which cannot be captured in a non-relativistic framework. We have used a CVS-4c-EOM-CCSD implementation for simulating the L-edge spectra of the Ar atom and compared our result with the experiment[48]. We have used the fully uncontracted version of 6-311(2+, +) G(p,d) with additional Rydberg functions as described by Vidal et al[47]. The peaks resulting from the transition from $^2P_{3/2}$ state are known as $L_{III}$-edge and those originating from $^2P_{1/2}$ are called $L_{II}$-edge. The XAS simulated and experimental L-edge XAS spectra of Ar atom[48] have been shown in Figure 2. The CVS-4c-EOM-CCSD result (blue line) is in good agreement with the experiment with just a shift of + 0.7 eV in the excitation energy. The first and second peak around ~244.5 eV and ~246.5 eV correspond to the transitions from $^2P_{3/2}(L_{III}) \to 4s$ and $^2P_{1/2}(L_{II}) \to 4s$, respectively. The third relatively small intense peak near 247 eV is due to the transition from $^2P_{3/2}(L_{III}) \to 5s, 3d$ and the small band arises at ~249eV, indicating a transition from $^2P_{3/2}(L_{II}) \to 5s, 3d$. The spectrum arising from the non-relativistic calculation (Red line) clearly shows a missing peak in the range 246-247 eV, which is correctly reproduced in the 4-c-EOM-CCSD method. It demonstrates the capability of the 4c-EOM-CCSD framework to simulate L-edge spectra of atoms with SOC-induced transitions accurately.

## 4. CONCLUSIONS

We present the theory, implementation, and benchmarking of transition properties within the framework of the four-component relativistic equation of motion coupled cluster method. The computational cost of calculating transition properties is at least twice that of the energy calculations. The 4c-EOM-CCSD method gives accurate excitation energy and transition dipole moments for systems containing heavy elements. The calculated transition moments show good agreement with the available experimental values and earlier reported theoretical results. The 4c-EOM-CCSD method can accurately simulate the low-intensity excited states arising due to the spin-orbit coupling. The 4c-EOM-CCSD can be extended to L-edge spectra using the CVS approximation. The simulated CVS-4c-EOM-CCSD spectrum agrees well with the experimental spectrum and correctly simulates the low-intensity spin-forbidden transition missing in the non-relativistic framework.

The present implementation will allow one to accurately simulate valence and core-excitation spectra of atoms and molecules using the relativistic equation of motion coupled cluster method. One can extend the applicability of the present implement to larger molecules using density fitting and natural spinor based approximation. Work is in progress towards that direction.

## ACKNOWLEDGMENTS

The authors acknowledge the support from the IIT Bombay, CRG and Matrix project of DST-SERB, CSIR-India, DST-Inspire Faculty Fellowship, Prime Minister's Research Fellowship, ISRO for financial support, IIT Bombay super computational facility, and C-DAC Supercomputing resources (PARAM Yuva-II, Param Bramha) for computational time.

**CONFLICT OF INTEREST**

The authors declare no competing financial interest.

**TABLE I.** Excitation Energy (EE in cm$^{-1}$) and Transition dipole moment (TDM in a.u.) for different excitations from the ground state (5p$^6$ ($^1S_0$)) of the Xe atom.

| Excited State | EOM-CCSD[a] | | EOM-CCSD[b] | | FSCCSD[c] | | EXP[d] | |
|---|---|---|---|---|---|---|---|---|
| | EE | TDM | EE | TDM | EE | TDM | EE | TDM |
| $5p^5(^2P_{3/2})6s\ ^2[3/2]_1^o$ | 67,145 | 0.648 | 67,886 | 0.647 | 68,147 | 0.634 | 68,045 | 0.654 |
| $5p^5(^2P_{1/2})6s\ ^2[1/2]_1^o$ | 76,554 | 0.537 | 77,166 | 0.535 | 77,201 | 0.512 | 77,185 | 0.521 |
| $5p^5(^2P_{3/2})5d\ ^2[1/2]_1^o$ | 80,487 | 0.012 | 80,368 | 0.058 | 80,259 | 0.101 | 79,987 | 0.120 |
| $5p^5(^2P_{3/2})5d\ ^2[3/2]_1^o$ | 87,638 | 0.931 | 86,328 | 0.949 | 84,137 | 0.663 | 83,889 | 0.704 |

[a] 4c-EOM-CCSD using d-aug-dyall.ae2z
[b] 4c-EOM-CCSD using d-aug-dyall.ae3z
[c] Fock space relativistic coupled cluster[16]
[d] Experiment[25]

**TABLE II. Excitation Energy (EE in cm$^{-1}$) and Transition Dipole Moment (TDM in a.u.) for different excitations from the ground state (X $^1\Sigma^+$) of HI molecule.**

| Excited State | 4c-EOM-CCSD[a] | | 4c-EOM-CCSD[b] | | MRD-CI[c] | | EXP[d] |
|---|---|---|---|---|---|---|---|
| | EE | TDM | EE | TDM | EE | TDM | EE |
| $a\ ^3\Pi_1$ | 40,510 | 0.0626 | 40,735 | 0.0753 | 40,220 | 0.0654 | 40,650 |
| $a\ ^3\Pi_{0+}$ | 44,426 | 0.1631 | 44,702 | 0.1644 | 44,007 | 0.1519 | 44,150 |
| $A\ ^1\Pi_1$ | 46,558 | 0.1343 | 46,775 | 0.1420 | 46,310 | 0.1538 | 48,150 |

[a] 4c-EOM-CCSD using d-aug-dyall.ae2z
[b] 4c-EOM-CCSD using d-aug-dyall.ae3z
[c] Relativistic MRD-CI[34]
[d] Experiment[36]

**TABLE III.** Spin-forbidden ($^1S_0 \to {}^3P_1$) and spin-allowed ($^1S_0 \to {}^1P_1$) transitions in closed-shell cations with excitation energy (in eV) and associated transition dipole moment (in a.u.), compared with previous theoretical[38] and experimental values[37].

| Method/transition | | Na$^+$ | | Mg$^{2+}$ | | K$^+$ | | Ca$^{2+}$ | |
|---|---|---|---|---|---|---|---|---|---|
| | | EE | TDM | EE | TDM | EE | TDM | EE | TDM |
| X2C-TD-EOM-CCSD[a] | $^1S_0 \to {}^3P_1$ | 35.1108 | | 54.8683 | | 21.2317 | | 31.0904 | |
| | $^1S_0 \to {}^1P_1$ | 35.3054 | | 55.2221 | | 21.4773 | | 31.4645 | |
| | Splitting | 0.1946 | | 0.3538 | | 0.2456 | | 0.3741 | |
| | MAE | | | | 1.4061 | | | | |
| 4c-EOM-CCSD[a*] | $^1S_0 \to {}^3P_1$ | 32.4071 | 0.061 | 52.4147 | 0.052 | 20.0505 | 0.197 | 29.9863 | 0.205 |
| | $^1S_0 \to {}^1P_1$ | 32.8124 | 0.275 | 53.0352 | 0.221 | 20.4246 | 0.512 | 30.4872 | 0.477 |
| | Splitting | 0.4053 | | 0.6205 | | 0.3741 | | 0.5009 | |
| | MAE | | | | 0.3629 | | | | |
| X2C-TD-EOM-CCSD[b] | $^1S_0 \to {}^3P_1$ | 32.9622 | | 53.0249 | | 20.2357 | | 30.2354 | |
| | $^1S_0 \to {}^1P_1$ | 33.2667 | | 53.5256 | | 20.5704 | | 30.6781 | |
| | Splitting | 0.3045 | | 0.5007 | | 0.3347 | | 0.4427 | |
| | MAE | | | | 0.0387 | | | | |
| TD-X2C-B3LYP[b] | $^1S_0 \to {}^3P_1$ | 30.1930 | | 49.4285 | | 19.6621 | | 28.7842 | |
| | $^1S_0 \to {}^1P_1$ | 30.5015 | | 49.9755 | | 19.8995 | | 29.2494 | |
| | Splitting | 0.3086 | | 0.5470 | | 0.2375 | | 0.4651 | |
| | MAE | | | | 2.2521 | | | | |
| TD-X2C-BP86[b] | $^1S_0 \to {}^3P_1$ | 29.2987 | | 48.8303 | | 18.8295 | | 28.7896 | |
| | $^1S_0 \to {}^1P_1$ | 29.5308 | | 49.2456 | | 19.1039 | | 29.2005 | |
| | Splitting | 0.2321 | | 0.4153 | | 0.2744 | | 0.4109 | |
| | MAE | | | | 2.7116 | | | | |
| TD-X2C-BHandH[b] | $^1S_0 \to {}^3P_1$ | 31.9982 | | 51.3506 | | 19.8055 | | 29.5461 | |
| | $^1S_0 \to {}^1P_1$ | 32.3611 | | 51.9756 | | 20.1467 | | 30.0574 | |
| | Splitting | 0.3629 | | 0.6250 | | 0.3412 | | 0.5113 | |
| | MAE | | | | 0.9101 | | | | |
| 4c-EOM-CCSD[b*] | $^1S_0 \to {}^3P_1$ | 33.0206 | 0.075 | 53.0320 | 0.062 | 20.2456 | 0.209 | 30.1706 | 0.203 |
| | $^1S_0 \to {}^1P_1$ | 33.3333 | 0.255 | 53.5486 | 0.217 | 20.5925 | 0.470 | 30.6593 | 0.437 |
| | Splitting | 0.3127 | | 0.5166 | | 0.3469 | | 0.4887 | |
| | MAE | | | | 0.0525 | | | | |
| Experiment | $^1S_0 \to {}^3P_1$ | 32.9413 | | 52.9249 | | 20.2382 | | 30.2435 | |
| | $^1S_0 \to {}^1P_1$ | 33.3224 | | 53.5029 | | 20.6381 | | 30.7104 | |
| | Splitting | 0.3811 | | 0.5780 | | 0.3999 | | 0.4669 | |

[a] 6-31G basis set is used for all the calculations
[a*] uncontracted 6-31G basis set is used for all the calculations
[b] Sapproro-DZP-2012-ALL basis set is used for all the calculations
[b*] uncontracted Sapproro-DZP-2012-ALL basis set is used for all the calculations

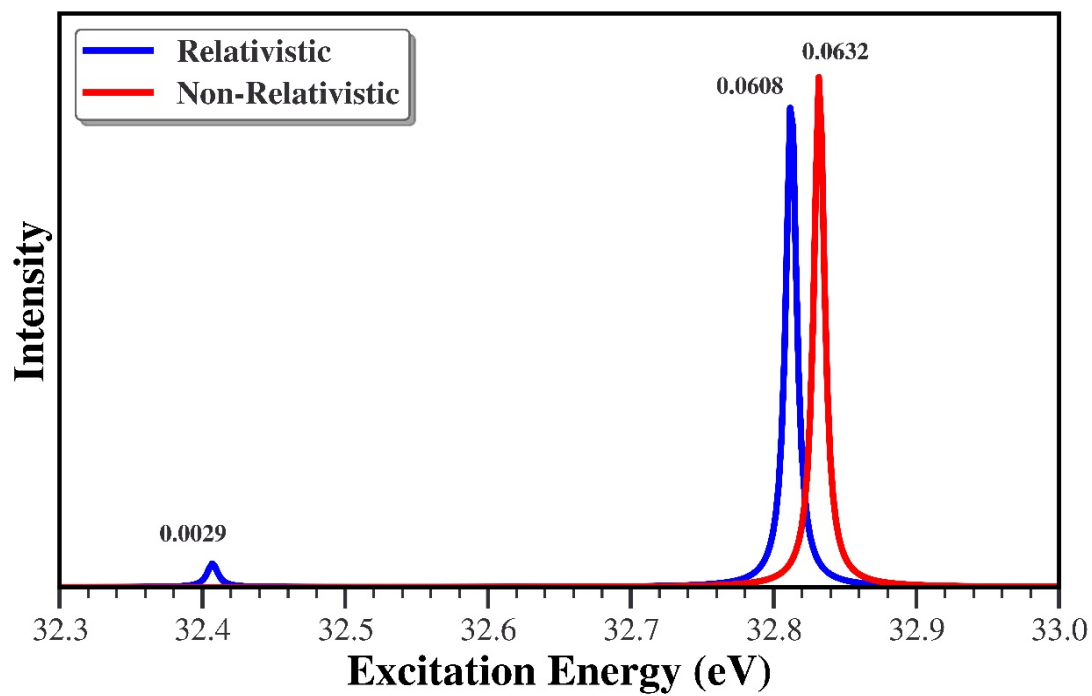

*FIG. 1. Absorption spectrum of Na⁺ atom calculated using four-component relativistic and non-relativistic EOM-CCSD method.*

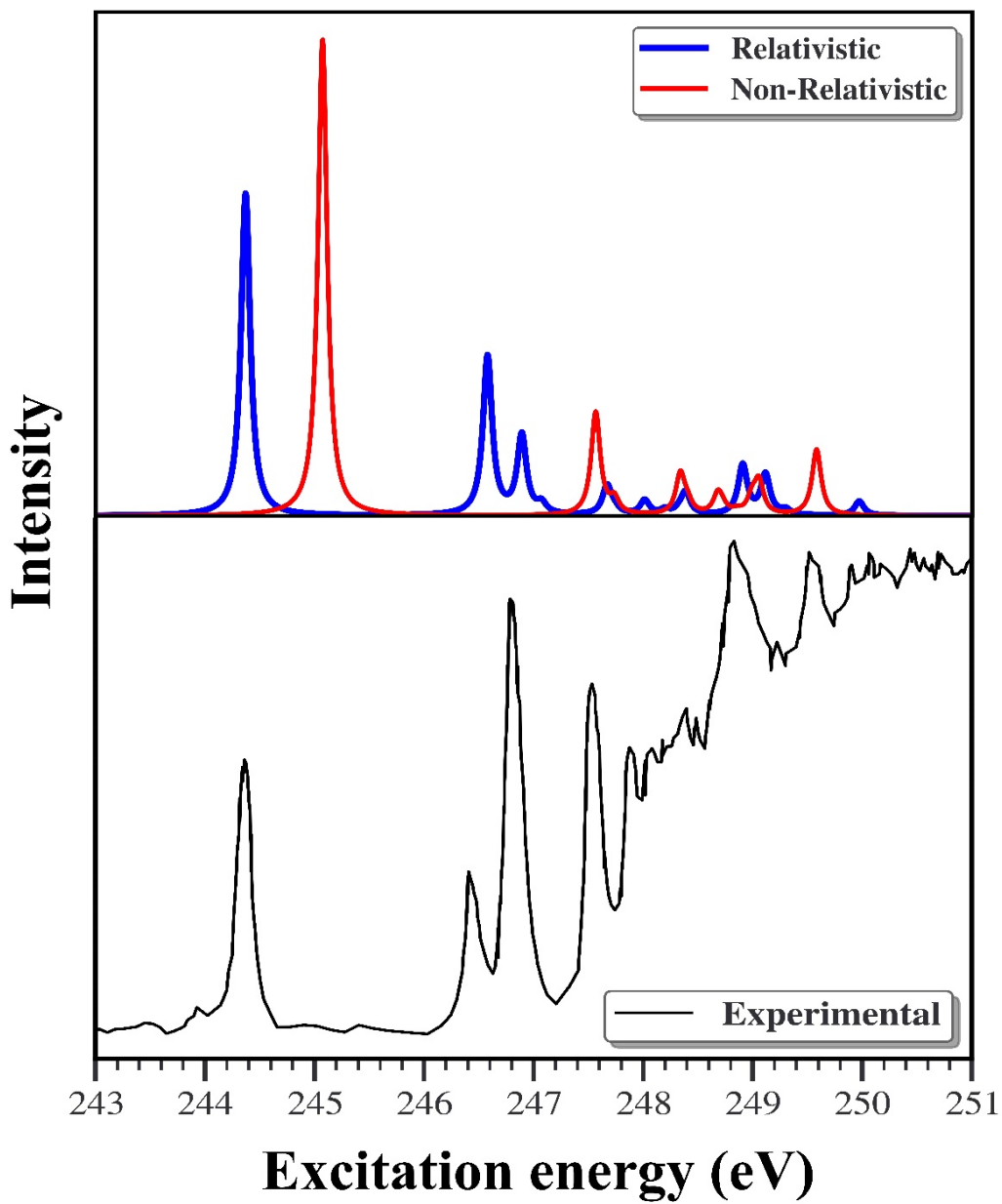

FIG. 2. L-edge X-ray absorption spectrum (XAS) of Ar atom calculated using relativistic CVS-4c-EOM-CCSD and non-relativistic CVS-EOM-CCSD method.